\documentclass{article}

\usepackage{amsmath, amssymb}
\usepackage{microtype}
\usepackage{graphicx}
\usepackage{subfigure}
\usepackage{booktabs} 

\usepackage[normalem]{ulem} 
\usepackage{xspace}
\usepackage{url}

\usepackage{breakurl}
\usepackage[breaklinks]{hyperref}
\usepackage{enumitem}

\setlist[itemize]{noitemsep, topsep=0pt}

\newcommand{\sys}{DeepPool\xspace}

\usepackage[accepted]{mlsys2022}

\mlsystitlerunning{Efficient Strong Scaling Through Burst Parallel Training}

\newcommand\redstrike[1]{\unskip}

\newcommand\greenstrike[1]{\unskip}

\newcommand\orangestrike[1]{\unskip}

\newcommand\bluestrike[1]{\unskip}
\newcommand\purple[1]{\unskip}
\newcommand\purplestrike[1]{\unskip}

\begin{document}

\twocolumn[
\mlsystitle{Efficient Strong Scaling Through Burst Parallel Training}



\mlsyssetsymbol{equal}{*}

\begin{mlsysauthorlist}
\mlsysauthor{Seo Jin Park}{csail}
\mlsysauthor{Joshua Fried}{csail}
\mlsysauthor{Sunghyun Kim}{csail}
\mlsysauthor{Mohammad Alizadeh}{csail}
\mlsysauthor{Adam Belay}{csail}
\end{mlsysauthorlist}

\mlsysaffiliation{csail}{MIT CSAIL}

\mlsyscorrespondingauthor{Seo Jin Park}{seojin@csail.mit.edu}

\mlsyskeywords{Distributed Machine Learning, MLSys}

\vskip 0.3in

\begin{abstract}
As emerging deep neural network (DNN) models continue to grow in size, using large GPU clusters to train DNNs is becoming an essential requirement to achieving acceptable training times.
In this paper, we consider the case where future increases in cluster size will cause the global batch size that can be used to train models to reach a fundamental limit: beyond a certain point, larger global batch sizes cause sample efficiency to degrade, increasing overall time to accuracy.
As a result, to achieve further improvements in training performance, we must instead consider ``strong scaling'' strategies that hold the global batch size constant and allocate smaller batches to each GPU. Unfortunately, this makes it significantly more difficult to use cluster resources efficiently.
We present \sys, a system that addresses this efficiency challenge through two key ideas. First, \emph{burst parallelism} allocates large numbers of GPUs to foreground jobs {\em in bursts} to exploit the unevenness in parallelism across layers. Second, \emph{GPU multiplexing} prioritizes throughput for foreground training jobs, while packing in background training jobs to reclaim underutilized GPU resources, thereby improving cluster-wide utilization. Together, these two ideas enable \sys to deliver a 1.2 -- 2.3$\times$ improvement in total cluster throughput over standard data parallelism with a single task when the cluster scale is large.
\end{abstract}
]



\printAffiliationsAndNotice{}  

\section{Introduction}

The size of deep learning models has been growing rapidly. For example, Microsoft announced that it successfully trained a model for machine translation with 17 billion parameters~\cite{t-nlg}. To train such large models quickly, it is necessary to use increasingly large training clusters. Current, state-of-the-art training systems already make use of hundreds of GPUs~\cite{goyal2017accurate, megatron-lm, MLSYS2021_28dd2c79}.

One of the most difficult challenges these systems face is maintaining efficiency while scaling. The conventional wisdom is to use data parallelism to distribute work across GPUs and to increase global batch sizes proportionally with the size of the cluster. This keeps the iteration time relatively constant, but allows for more samples to be processed per iteration, increasing training throughput with cluster size.

Unfortunately, a higher sample processing throughput does not necessarily translate to a faster time to accuracy: beyond a certain point, training algorithms become less effective with larger batch sizes because they experience a loss in sample efficiency~\cite{shallue2018measuring, pollux}. Ultimately, this places a fundamental limit on the maximum global batch size that can be beneficial. Therefore, as the number of GPUs used in training continues to increase, each GPU will inevitably be forced to process smaller batches at a time.

In this paper, we investigate how to balance cluster utilization with time to accuracy in the regime where batch sizes are limited by sample efficiency. The key difficulty is that smaller per-GPU batch sizes expose the uneven parallelism inherent to many DNN models: some layers have sufficient parallelism to keep GPUs busy, while others leave GPUs significantly underutilized. As a result, a conventional data parallel approach is insufficient to make efficient use of GPU hardware.

To address this challenge, we propose \textbf{\sys{}}, a system that realizes two key ideas. First, we introduce \emph{burst parallel training} to dynamically adjust the number of GPUs allocated to each layer, so that layers with less parallelism can use fewer GPUs. This increases overall cluster efficiency because it frees up underutilized GPUs for use by other training tasks. Second, we propose a new collection of \emph{GPU multiplexing} techniques that allow multiple models to be trained on each GPU simultaneously. Intuitively, this allows us to further improve efficiency because GPUs have massive internal parallelism that is underutilized by smaller per-GPU batches (SMs, tensor cores, memory bandwidth, etc.). Thus, collocating training tasks can yield better GPU throughput overall.

Throughout this paper, we distinguish between foreground training tasks, such as the training of production models, and background training tasks, like testing new model designs at a smaller scale. \sys{} tailors both of its mechanisms to minimize time to accuracy for foreground tasks, while improving overall cluster throughput by packing background tasks on to the same GPUs.

Our evaluation shows that \sys{} can achieve 1.2--2.3x improvements in total cluster throughput over standard data parallelism with a single task. 
When compared to statically partitioning a cluster into two groups (data-parallel foreground task training and individual background task training), \sys{} allows 11--38\% higher foreground speedup for the same high level of cluster throughput.

\sys{} has some limitations. First, it focuses only on dynamic scaling on sample dimensions, but a more general system would explore more flexible forms of model parallelism. Second, it currently limits background jobs to a single GPU, but our burst parallel planner could be extended in the future to scale background jobs across multiple GPUs. Finally, we discovered that today's GPU hardware has shortcomings that limit its multiplexing efficiency; we discuss our workarounds for this problem in \S\ref{sec:multiplexing}.
\section{Challenges for Scaling DNN Training Efficiently}
\label{sec:challenge}

There is significant interest in speeding up DNN training on large GPU clusters \cite{dean2012large, abadi2016tensorflow, jia2018data, granularStrongScale, ben2019demystifying, MLSYS2021_28dd2c79}. A primary metric for many organizations is the {\em time to accuracy}, i.e. how long it takes to train a model to a desired level of accuracy. The ML design process is often iterative, with researchers repeatedly training models and refining the architecture, training data, or hyperparameters based on outcomes. Improving time to accuracy for such jobs, which we will henceforth refer to as ``foreground'' jobs, can therefore significantly impact researcher productivity. 

Unfortunately, parallelizing a DNN training job on a large scale often makes inefficient use of GPU hardware. The conventional approach is to use {\em weak scaling}: as we scale a job to a large number of GPUs, we increase the global batch size correspondingly, such that the per-GPU batch size is kept constant. Weak scaling improves training {\em throughput} (the number of data samples processed per second) linearly with cluster size. However, it usually does not result in a linear reduction in time to accuracy. The reason is that increasing the global batch size beyond a certain point can degrade the sample efficiency of the learning algorithm (i.e., we require more data samples to reach the same level of accuracy). This drop in sample efficiency  results in a diminishing return in terms of wall-clock training time~\cite{shallue2018measuring}. At large scale, such as 128 GPUs or greater, the training time may entirely cease to improve with the addition of more GPUs.

\begin{figure}
\centering
\includegraphics[width=0.85\columnwidth]{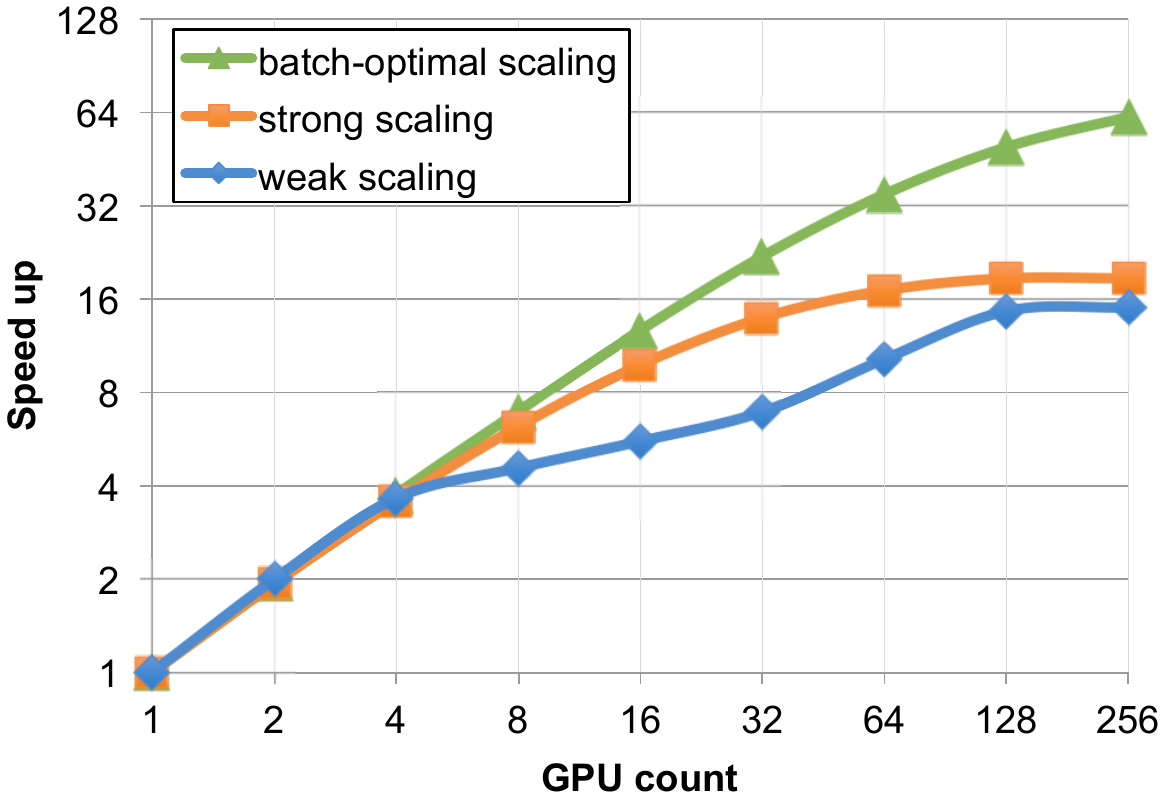}
        \caption{The estimated speedups for training VGG-11 to error = 0.35 with different scaling strategies. Each GPU has 1 Tbps full bi-section networking. Weak scaling uses 256 samples per GPU at each iteration. Strong scaling splits 256 samples across GPUs.}
\label{fig:strongScalingNoComm}
\end{figure}

An alternative approach for accelerating DNN training is {\em strong scaling}: given more GPUs,  keep the global batch size fixed, but reduce the amount of computation done by each GPU. For example, in data-parallel training, each GPU processes a smaller batch of samples in each iteration as we scale to more nodes. Unlike weak scaling, this approach does not degrade sample efficiency as it does not change the global batch size. However, it introduces two forms of inefficiency that again prevent a linear speedup with cluster size. First, since strong scaling reduces the amount of computation on each GPU in an iteration, the training job eventually gets bottlenecked by communication bandwidth between GPUs (e.g., for parameter synchronization in data-parallel training). Second, GPUs have massive internal parallelism, which is hard to utilize fully when the GPU's workload becomes too small (e.g., small batches or operations). 

An obvious improvement to these two extremes is to optimize the global batch size for each scale. This approach, which we will refer to as ``batch-optimal scaling'', seeks to find the sweet spot in training throughput and sample efficiency that provides the best time to accuracy.

Figure~\ref{fig:strongScalingNoComm} compares the estimated speedup achieved by the above three approaches for training the VGG-11 model. To obtain these estimates, we refer to the study of Shallue et al.~\cite{shallue2018measuring} to  determine the number of iterations required to train the model to desired accuracy at different global batch sizes. Next, we profile the computation time for different layers of VGG-11 on an NVIDIA A100 GPU at different batch sizes and use a simple network model to estimate the communication time for data-parallel training (see ``modeling communication cost'' in \S\ref{sec:burstParallel} for details), which together give an estimate of the time per iteration for each approach. As the figure shows, all approaches provide linear speedup up to 4 GPUs. Beyond this point, the marginal benefits of adding more GPUs diminish for weak scaling while strong and batch-optimal scaling provide better speedups. 

\begin{figure}
\centering
\includegraphics[width=0.85\columnwidth]{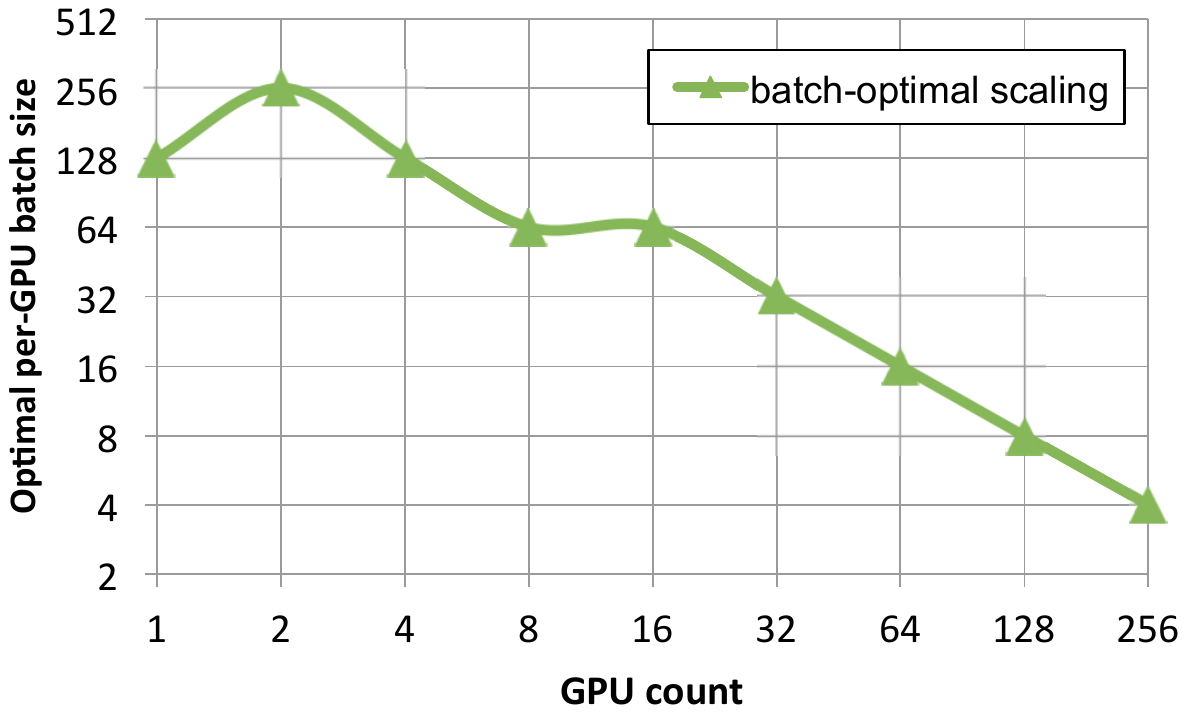}
        \caption{The per-GPU batch size for training VGG-11 to error = 0.35 chosen by the batch-optimal scaling strategy at each scale. Each GPU (NVIDIA A100) has 4.8 Tbps bi-directional networking. As the job scales, the per-GPU batch size tends to decrease, implying a lower per-GPU compute load. }
\label{fig:batchOptimalperGPUsize}
\end{figure}

This simple experiment reveals several interesting findings.

\noindent{\bf 1. Optimizing time-to-accuracy requires small per-GPU batches at large scale}. Figure~\ref{fig:batchOptimalperGPUsize} shows the per-GPU batch size chosen by the batch-optimal scaling strategy at each scale. As we can see, at small scale, the best strategy is to use a large per-GPU batch size. However, as the job scales, the optimal per-GPU batch size gets smaller, implying that the amount of compute performed by a GPU in an iteration must decrease. 

\begin{figure}
\centering
\includegraphics[width=0.85\columnwidth]{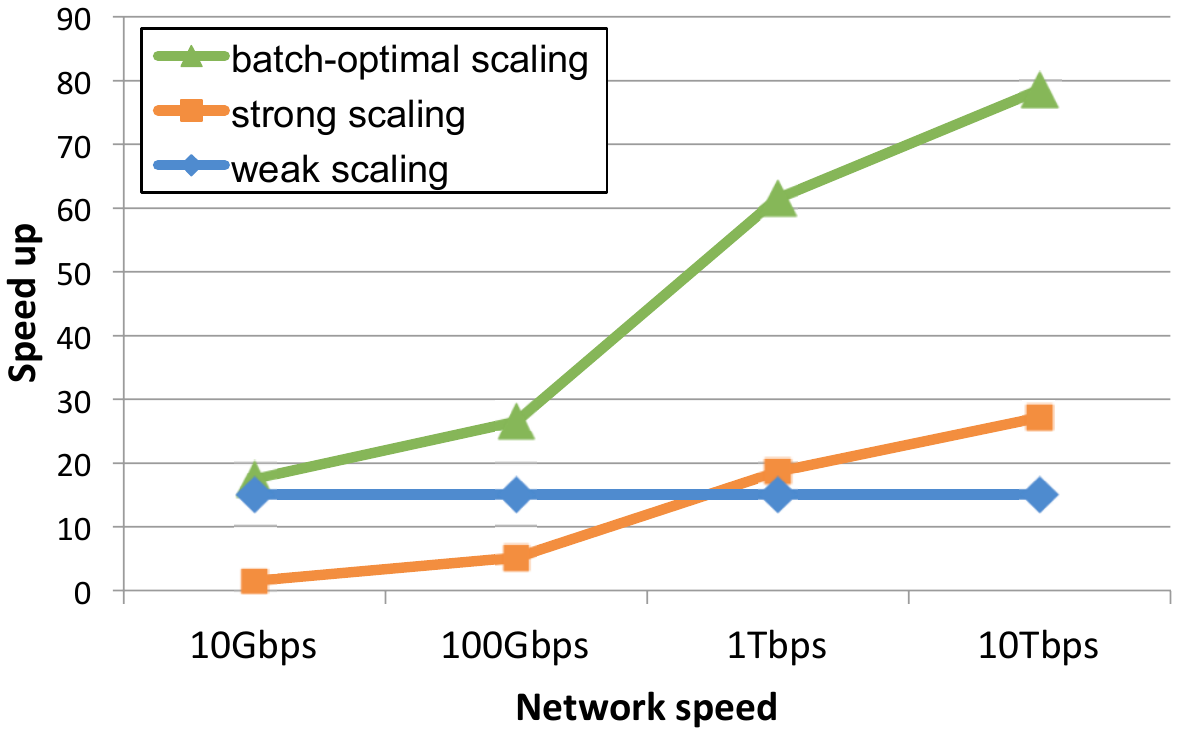}
        \caption{The estimated speedups by using 256 GPUs for training VGG-11 to error = 0.35 at four different network speeds. Weak scaling uses 256 samples per GPU, strong scaling splits 256 samples across GPUs. The benefits of using higher network speeds are clear in the strategies that employ strong scaling.}
\label{fig:optimalBatchByBetBW}
\end{figure}

\noindent{\bf 2. Strong scaling and small per-GPU batches are more effective with fast networks.} Figure~\ref{fig:optimalBatchByBetBW} compares the speedup of each approach at different network speeds at a scale of 256 GPUs. When the network is slow (e.g., 10~Gbps), weak scaling is preferrable since it amortizes the communication cost of parameter synchronization over a larger amount of computation in each iteration. However, with today's cutting edge networks (e.g., 2 Tbps NVSwitch, 200 Gbps ConnectX-6) and even faster network technologies on the horizon~\cite{sipml}, strong scaling will become more attractive. 

\begin{figure}
\centering
\includegraphics[width=\columnwidth]{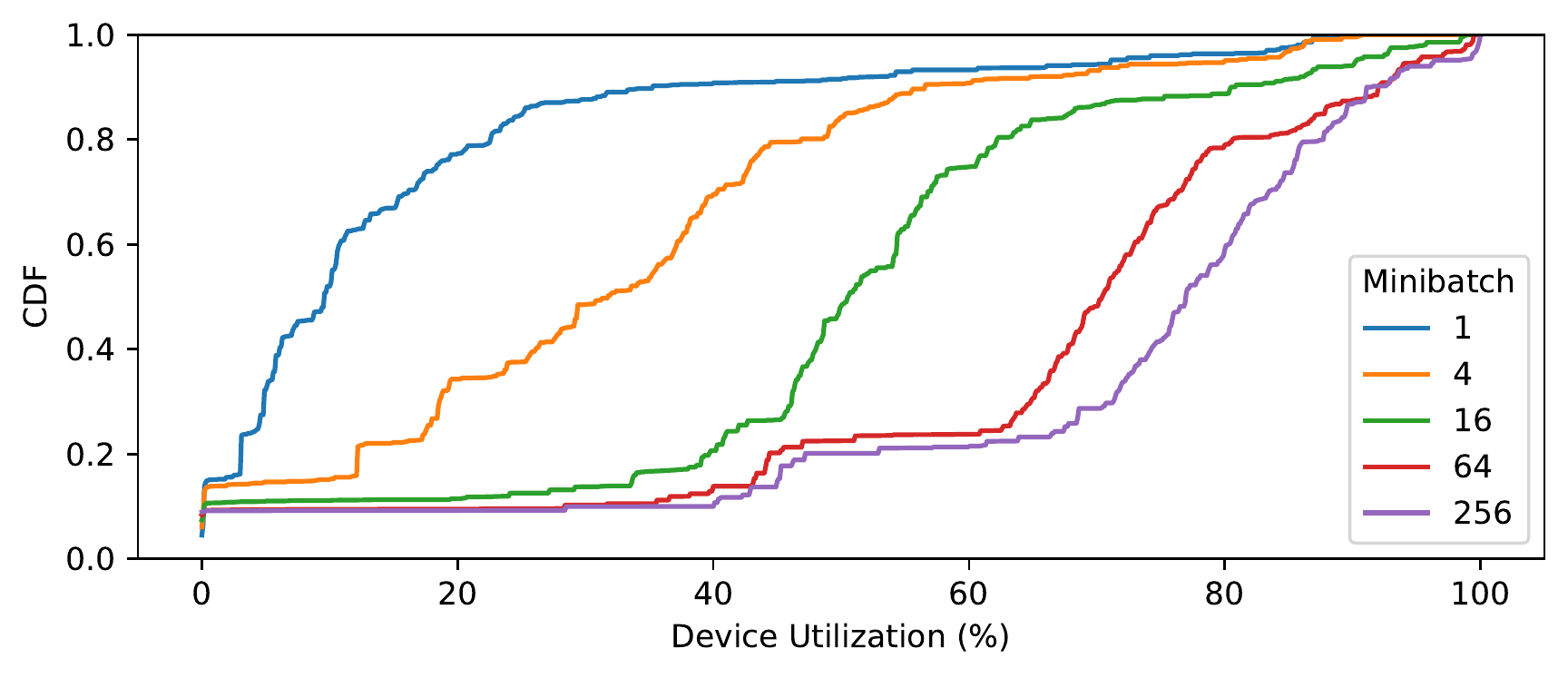}
\caption{GPU utilization of Resnet-50.}
\label{fig:gpuUnderutilization-resnet50}
\end{figure}

\noindent{\bf 3. None of the approaches achieve perfect linear scaling.} As shown in Figure~\ref{fig:strongScalingNoComm}, even the batch-optimal approach still suffers from diminishing and sub-linear speedup at large scale. As mentioned earlier, this is due to both communication overheads and GPU underutilization as we scale the cluster. Figure~\ref{fig:gpuUnderutilization-resnet50} shows the device utilization achieved at different batch sizes. For small batch sizes, GPU utilization degrades, suggesting that even with very fast networks, we cannot achieve a linear speedup. 

These findings suggest that operators of today's GPU clusters have to make an unfortunate choice. For an important foreground training job, they must either limit the scale, therefore completing the job slower than theoretically possible on the cluster. Alternatively, they can run the job at large scale and suffer low GPU utilization. However, since large GPU clusters are shared by multiple jobs, a natural question is: {\em is it possible to achieve speedups near the theoretical limits for foreground jobs while simultaneously achieving high GPU utilization by reclaiming unused cycles for less-time-critical background jobs?} 

In this paper, we present a system, \sys, to achieve precisely this --- strong scaling of foreground jobs with efficient GPU multiplexing between foreground and background jobs.

\section{Approach and Overview}

\subsection{Approach}

Our main tool to achieve efficient strong scaling is collocating more than one job on a single GPU. Large GPU clusters are typically shared by many users and training jobs (e.g., within an institution), and there are many small training jobs that don't require large-scale training (usually for quick testing of new model architecture with a small dataset or when low cost matters more than speed). By scheduling those small training jobs as a low-priority job, we can reclaim the spare compute power of GPUs during strong scaling. GPU's limited memory is often an obstacle to scheduling multiple jobs on a GPU. Fortunately, strong scaling reduces memory footprints as well as compute load, reserving enough memory space for a small background job.

\begin{figure}
\centering
\includegraphics[width=\columnwidth]{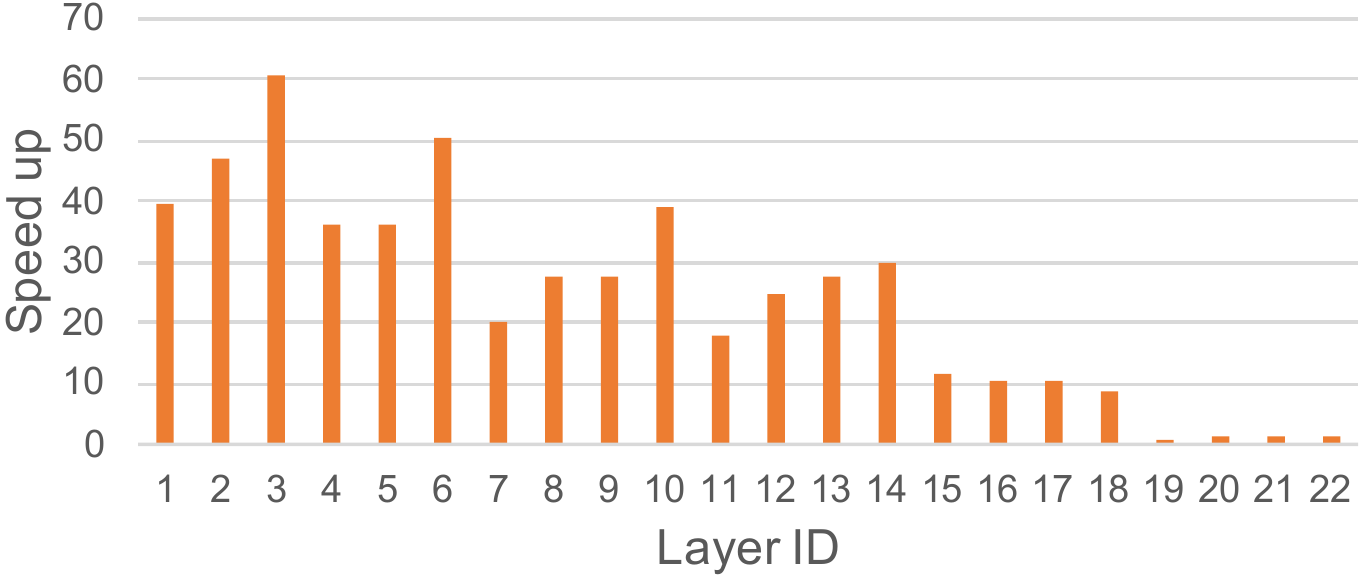}
\caption{Heterogeneous scalability of layers in VGG16. Y-axis shows the speedup of each layer when it is strong scaled from 128 samples per iteration to 2 samples per iteration using 64 GPUs.}
\label{fig:heterogeneousScalability-vgg16}
\vspace{-1em}
\end{figure}

With the ability to reclaim underutilized GPUs with background jobs, we further improve efficiency with \textbf{burst parallel training}. One source of inefficiency in strong scaling is the uneven parallelism across stages within an iteration. As an example, Figure~\ref{fig:heterogeneousScalability-vgg16} shows the scalability of layers in VGG16; some layers can achieve near-linear speedup with more GPUs, but some other layers don't get any faster. By bursting the number of GPUs only for training stages that can benefit from more GPUs, we can drive up the efficiency further. When some GPUs are idling for the foreground job (for stages using a smaller number of GPUs), the background job can fully utilize the GPU. Alternatively, we may carefully place another large-scale distributed job during the idle gaps.

\subsection{System overview}

\begin{figure}
\centering
\includegraphics[width=0.9\columnwidth]{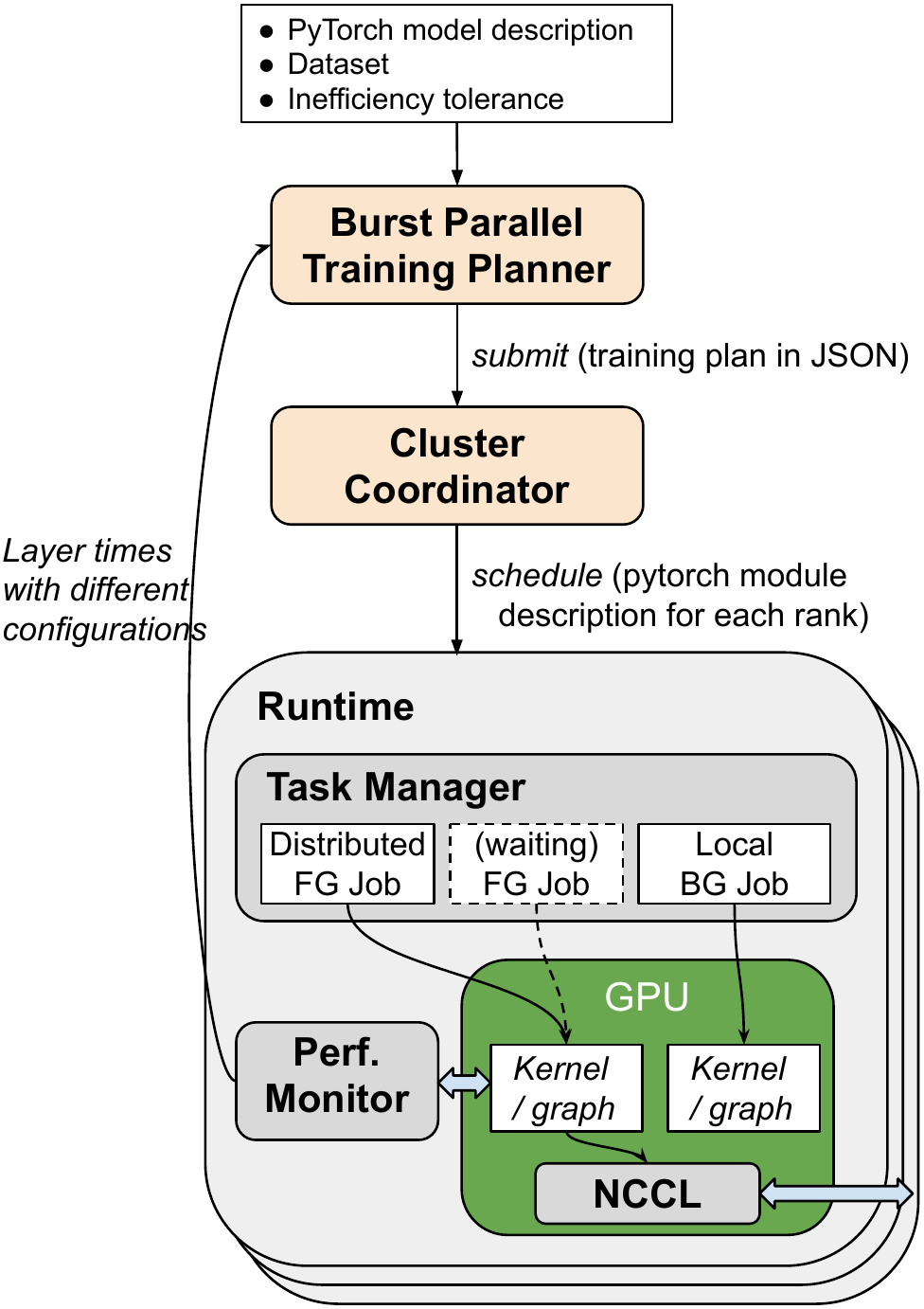}
        \caption{System architecture.}
\label{fig:architecture}
\end{figure}

We built \sys , a prototype training system that achieves efficient strong scaling by implementing burst parallel training and GPU multiplexing. 

A user may submit a training job to \sys by providing a PyTorch-like model implementation, dataset, and \emph{inefficiency tolerance}. With the limit on scaling inefficiency, the \textbf{burst parallel training planner} decides the scaling of each layer so that it won't use too many GPUs and cause inefficiency beyond the allowance. For estimating speedup and inefficiency, the planner initially profiles each layer with different batch sizes. 
For the planner to work properly, DeepPool requires the input model's execution graph to be static.

The generated burst parallel training plan is then handed to the cluster coordinator, which manages all GPU runtimes and places a new training job on a subset of GPUs, as specified in the training plan. Currently, the cluster coordinator doesn't support complex placement such as aligning gaps of one burst parallel job with another burst parallel job.

Each GPU in the cluster has a separate host-run \sys \emph{runtime}. The \emph{task manager} in each runtime manages all training jobs including the background job. At any moment, a task manager schedules one distributed foreground (burst parallel) job and one local low-priority background job, and only the foreground job communicates with other GPUs via NVIDIA Collective Communications Library (NCCL). The runtime uses the C++ frontend of PyTorch \cite{libtorch} as the execution engine. While running training jobs, the performance monitor module profiles the runtime of each layer, which may be fed back to the planner to optimize the training plan. In our prototype, this feedback loop happens manually. Additionally, the profiling data is used to bound any slowdowns that may be caused by poor collocation choices.

In the following sections, we discuss how the burst-parallel training planner optimizes the parallelization plan (\S\ref{sec:burstParallel}) and how we minimize the interference between foreground jobs and background jobs (\S\ref{sec:multiplexing}).

\section{Burst Parallel Scheduling}
\label{sec:burstParallel}
\sys{}'s burst parallel training planner takes as input the total number of available GPUs and the global batch size, which may be optimized by statistical efficiency analysis as seen in \S\ref{sec:challenge}.
Given the two parameters, the planner finds the best strong scaling strategy by choosing the level of scaling (the number of GPUs) for every layer in the model. It minimizes iteration time while limiting the scale of each stage up to a user-given efficiency limit.

Scaling efficiency is defined by \textit{GPU-sec} amplification, where \textit{GPU-sec} is the aggregate active GPU time per iteration, like man-hour or watt-hour. Amplification is defined as $\langle\textit{GPU-sec when scaled}\rangle / \langle\textit{single GPU iteration time}\rangle$.
The planner considers the GPU-sec amplification of each layer individually.

\subsection{Iteration time components}
\sys{}'s planner estimates iteration time by summing up each layer's computation and communication costs. 

\textbf{Computation cost.}
To compute the iteration time, the planner profiles the computation costs of each layer with every possible degree of scaling. We measure each layer's time to compute forward and backward passes with different per-GPU batch sizes, and the sum is used. In formal notation,

\begin{itemize}
\item \texttt{comp$_{(i,g)}$}: the sum of forward and backward compute time of $i$-th layer when scaled to $g$ GPUs.
\end{itemize}

\textbf{Communication cost.}
We use a simple networking model for estimating communication costs during training. We assume full bi-section networking (as in NVSwitch networking) and profile the networking performance in terms of per-GPU bandwidth and minimum propagation delay. In the planner, we simply divide the payload size by the bandwidth and add the propagation delay. 

The planner must also account for the communication involved when scaling the number of GPUs up or down between layers. When this happens, samples must be transferred across GPUs before running the layer (as do gradients during backward passes). In formal notation,

\begin{itemize}
\item \texttt{comm$_{(i,g) \rightarrow (j,h)}$}: communication time for activation and back-propagation between $i$-th layer and $j$-th layer when scaled to $g$ GPUs and $h$ GPUs respectively.
\end {itemize}

Another communication overhead considered is gradient/parameter synchronization after a backward pass. For simplicity, we assume that this synchronization does not overlap with the backward pass. In formal notation,

\begin{itemize}
\item \texttt{sync$_{(i,g)}$}: time for synchronizing gradients of $i$-th layer when scaled to $g$ GPUs.
\end{itemize}

\subsection{Search algorithm}
Naively bursting individual layers to the amplification limit won't give the best iteration time since frequent changes of scaling lead to higher communication costs. Thus, \sys{} needs a search algorithm to find the optimal parallelization plan.

The parallelization strategy search is composed of two parts. We use a dynamic programming algorithm that finds the optimal plan for a single chain of DNN layers, and we perform a graph reduction algorithm for DNN graphs that branch out to multiple parallel chains of layers.

\textbf{Linear search on a single chain of layers.} Our dynamic programming for a chain of layers is similar to shortest path search except that we consider the user-given limit on GPU-sec amplification. As input for search, we are given a chain of $L$ layers indexed by $[1, L]$ and the number of available GPUs, $G$.

With dynamic programming, we compute two tables:

    $ S[i][g] = \textrm{ shortest time to complete } L_1, \dots, L_i $
    
    \hspace{6em} $\textrm{ while using } g \textrm { GPUs for } L_i.$

    $T[i][g] = \textrm{time spent on } L_i \textrm{ while minimizing } S[i][g]$

\begin{algorithm}[tb]
   \caption{Search Linear}
   \label{alg:linearSearch}
\begin{algorithmic}   
   \STATE Initialize $S[0][g]=0$ for all $g \in [1, G]$
   \FOR{$i=1$ {\bfseries to} $L$}
   \FOR{$g$ {\bfseries in} $[1, G]$}
   \STATE Initialize bestAmp = $\infty$.
   \STATE Initialize bestS = $\infty$.
   \STATE Initialize bestT = $\infty$.
   \FOR{$h$ {\bfseries in} $[1, G]$}
   
   \IF{Amp($i-1, h$) $\le$ max(bestAmp, AmpLimit) {\bfseries and} 
        $S[i-1][h] + \texttt{comm}_{(i-1,h) \rightarrow (i,g)} \le$  bestS}
   
   \STATE bestS $\leftarrow S[i-1][h] + \texttt{comm}_{(i-1,h) \rightarrow (i,g)}$
   \STATE bestT $\leftarrow \textrm{\texttt{comm}}_{(i-1,h) \rightarrow (i,g)}$
   \STATE bestAmp $\leftarrow$ min(bestAmp, Amp($i-1, h$))

   \ENDIF
   \ENDFOR
    \STATE $S[i][g] =$ bestS + { \texttt{comp}$_{(i,g)}$} + \texttt{sync}$_{(i,g)}$
    \STATE $T[i][g] =$ bestT + { \texttt{comp}$_{(i,g)}$} + \texttt{sync}$_{(i,g)}$
   \ENDFOR
   \ENDFOR
\end{algorithmic}
\end{algorithm}

\begin{figure*}
\centering
\includegraphics[width=0.8\textwidth]{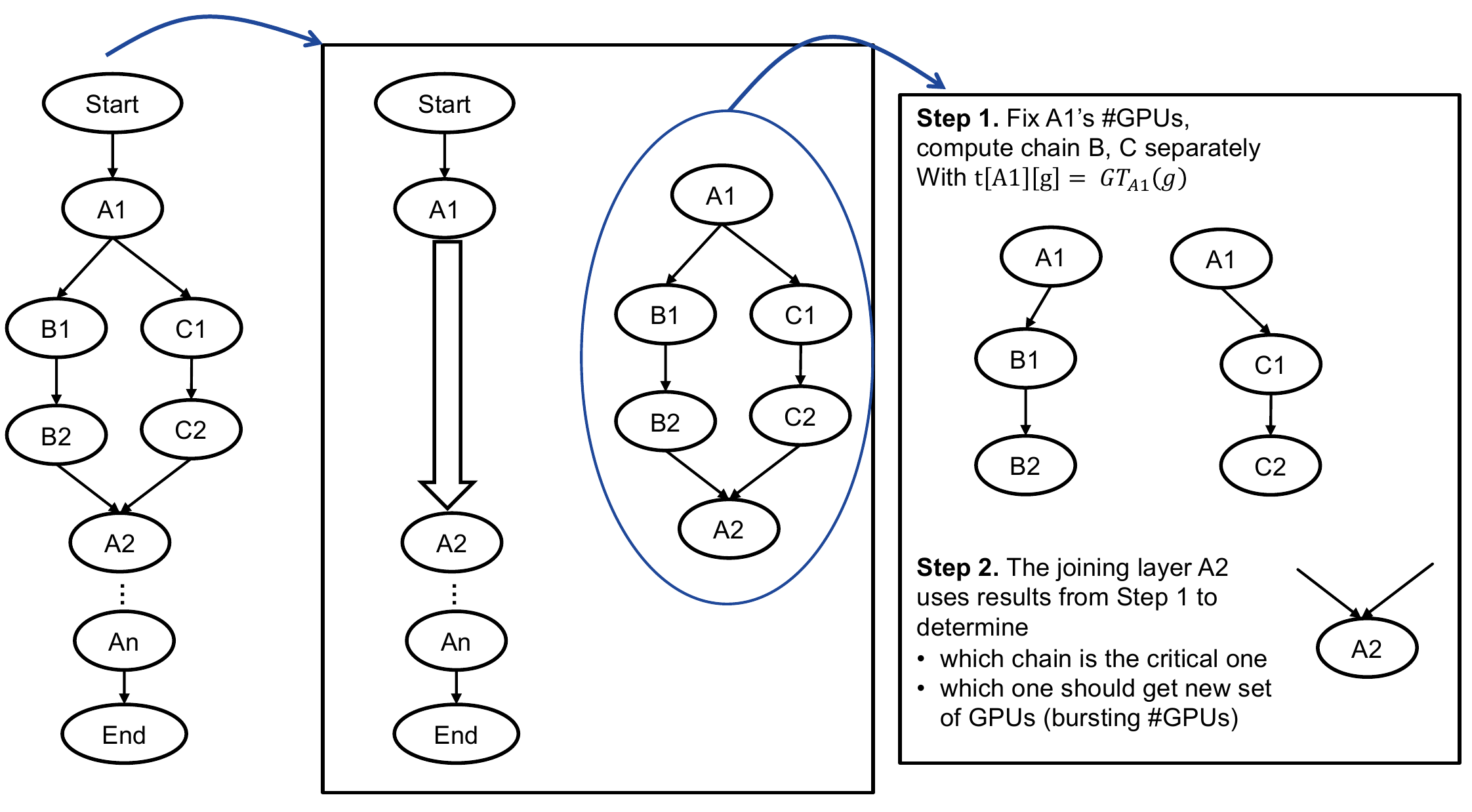}
\caption{The graph reduction process of multi-chain search problem into a single chain search problem.}
\label{fig:decomposingSearch} 
\end{figure*}

\vspace{1em}
The dynamic programming proceeds as shown in Algorithm~\ref{alg:linearSearch}. In the algorithm description, $Amp (i, g)$ denotes the GPU-sec amplification for $L_i$ when scaled to $g$ GPUs.

\[ { Amp (i, g)} =  \frac{T[i][g] \times g}{ \texttt{comp}_{(i,1)} }\]

Note that $T[i][g]$ includes communication overheads (\texttt{comm} and \texttt{sync}), so the amplification limit also counts the communication in. After populating $S$, we can backtrace from $L_L$ to $L_1$ and find the number of GPUs used by each layer that minimize the total processing time while obeying the amplification limit.

\textbf{Multi-chain graph reduction.} We extend the single chain algorithm to an algorithm for general DNN computation graphs with branches and joins. The algorithm identifies portions of DNN graphs from the branching layer to the joining layer, and multiple chains inside each block are reduced to a single edge. Figure~\ref{fig:decomposingSearch} shows how we reduce a complex graph into a chain of blocks, where each block branches at the beginning and joins at the end. When the linear search algorithm hits a branching layer (A1 in the example), it finds the corresponding joining layer (A2). The linear search algorithm consider the whole block as a single edge cost. 

\begin{itemize}
\item $\texttt{tr}_{(i,g) \rightarrow (j,h)}$: time elapsed to transit from $L_i$ to $L_j$ when scaled to $g$ and $h$ GPUs respectively. This is same as \texttt{comm} if $L_i$ is directly followed by  $L_j$.
\end{itemize}

Given that we can reduce branch-join blocks as the transition cost, the linear search algorithm is identical to the single-chain search algorithm.

To compute the transition time between the branching layer and the joining layer, we utilize linear searches on branches. First, we fix the number of GPUs for the branching layer with one of the possible numbers, $g$. Then, we perform linear searches on all branches while restricting A1 must be scaled to $g$. After the individual linear search on each branch, we merge the shortest time to reach the joining layer. The joining layer figures out the critical branch which takes the longest time and decides whether or not to run each other non-critical branch in parallel with the critical branch. For that consideration, we use communication overheads to move samples to and from a new set of GPUs and make sure it won't increase the total training time or overshoot the amplification limit. After merging the shortest times of all branches for all possible GPU counts for the joining layer, we repeat the process with other GPU counts for the branching layer.

\section{Multiplexing}
\label{sec:multiplexing}
\begin{figure*}
\vspace{-0.5em}
\centering
\includegraphics[width=0.8\textwidth]{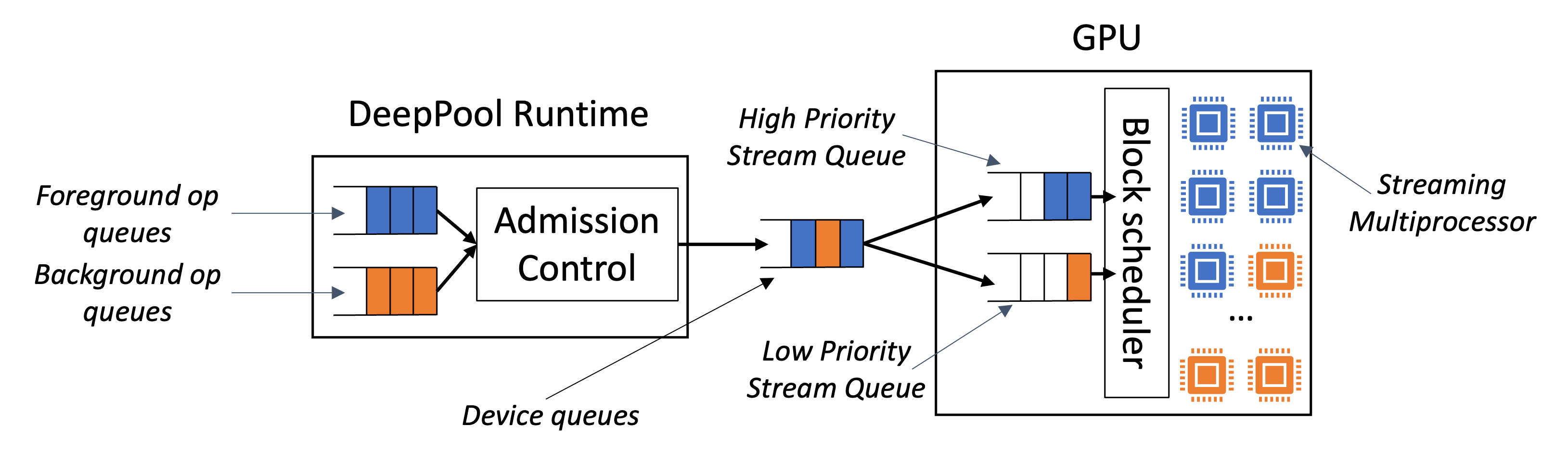}
\caption{Overview of queues in \sys{}. DeepPool queues work from high and low priority tasks, and uses CUDA stream priorities to provide QoS for high-priority tasks. Because some queues inside the CUDA driver are shared by high and low-priority work, \sys{} keeps queueing at the device to a minimum. }
\vspace{-1em}
\label{fig:systemqueues} 
\end{figure*}

\sys{} aims to make use of any unused compute resources left on each GPU in the cluster by concurrently executing a low-priority background job on the GPU alongside the main distributed training task. The goal is to maximize the throughput of the background job, raising the overall throughput and efficiency of the cluster, while ensuring that foreground jobs have nearly the same quality-of-service (QoS) that they would have if executing in isolation. Because it is one of the most common platforms used for training DNNs, \sys{} targets Nvidia's datacenter-class Tesla GPUs. While Nvidia's CUDA implementation provides some functionality for multiplexing tasks with different priorities, we find that the mechanisms are somewhat lacking. As a result, \sys{} must at times intentionally under-utilize the GPUs in order to protect QoS for the foreground job. In this section, we discuss the shortcomings of current GPU hardware for multiplexing, and the mechanisms that \sys{} leverages in order to use multiplexing to improve cluster utilization and efficiency, despite these shortcomings.

We start with some background on the execution model for CUDA devices. CUDA allows concurrent execution of tasks on the GPU using a mechanism called \emph{CUDA streams}. CUDA streams allow a program to express dependencies between various operations that are launched on the device. Operations (e.g. compute kernels and memory operations) that are enqueued into a stream are executed on the GPU in FIFO order, while operations placed into different streams may execute concurrently, assuming sufficient compute resources are available on the GPU's streaming multiprocessors (SMs). CUDA streams can be assigned an integer priority at creation time, which allows the on-device scheduler to favor scheduling operations from certain streams over others. Unfortunately, the on-device scheduler on Nvidia GPUs is non-preemptive: while it may favor running kernels from high-priority streams, once it assigns a block of computation to one of its SMs, it allows it to run to completion without interrupting it. The limitation impacts our ability to extract higher utilization from the device when multiplexing high-priority and low-priority jobs: in order to reduce delays for the foreground job, \sys{} runs background DNN jobs with small batch sizes in order to decrease their execution latencies. Unfortunately, this also reduces the total possible utilization for the background job. 

In addition to the non-preemptive scheduler, we found several other idiosyncrasies in Nvidia's implementation of stream priorities. We found that work can queue in a number of places that do not differentiate among the respective priorities of the queued work. For example, we found that allowing an unbounded number of kernel or graph launches from each task leads quickly to loss of QoS for the foreground job. We speculate that this is because launch requests from different priority streams are placed into shared transmission queues between the CUDA driver and the device, potentially starving the device of high-priority requests. \sys{} limits the number of outstanding queued requests to limit the amount of head-of-line blocking possibly experienced by foreground jobs. Figure~\ref{fig:systemqueues} shows the various hardware queues in the system and illustrates our approach to managing them. We find that overall, several improvements could be made to Nvidia's software and hardware stack to improve efficiency, the most impactful being the addition of an on-device preemptive scheduler.

\sys{}'s runtime contains a lightweight execution engine that mediates kernel launches on the GPU from among the set of jobs that are currently assigned to that GPU. The engine maintains a virtual execution queue for each DNN task along with an associated CUDA stream, and limits the number of outstanding launches from each queue onto the GPU to avoid interference in the device's execution queues. The engine leverages \emph{CUDA graphs}, a recent CUDA feature addition that allows the program to group multiple consecutive operations together into a single launch operation, amortizing the costs of communicating with the device. Traditionally, CUDA programs submit each operation to the GPU using calls such as \texttt{cudaLaunchKernel}, which asynchronously enqueue the operation onto the device and return within several microseconds. When a computation contains small kernels with short execution times, the host can easily underutilize the device when the kernel runtime is shorter than the amount of time it takes for the host to prepare and enqueue the next operation. Many DNN models benefit heavily from CUDA graphs, particularly when the per-GPU minibatch size is low, and \sys{}'s execution engine enables it for all jobs that it runs.

In addition to improving task throughput and GPU utilization, CUDA graphs play an important role in maintaining good QoS for foreground jobs in \sys. Because the GPU scheduler is non-preemptive, if the host falls behind in supplying high-priority operations to the device to execute, the device will schedule a lower-priority operation leading to a potentially long delay once the host does enqueue the next operation. CUDA graphs reduce interference by reducing opportunities for low-priority tasks to use device cycles that could be more immediately used for high-priority jobs. While CUDA graphs largely improve the performance of \sys{}, we observe that the graph execution engine on the device is also vulnerable to head-of-line blocking for high-priority tasks when low-priority tasks have large graphs. To ensure that low-priority tasks with more kernels do not starve high-priority tasks, we split large CUDA graphs launches into groups of smaller graphs, and limit the number of outstanding graph launches. 

We also observed that a few training operations were particularly sensitive to interference from low priority tasks. For example, NCCL's all-reduce operation, which is used to synchronize gradients during the backwards pass, more than doubles in execution time when another task is run on the same GPU. To handle such cases, the execution engine monitors the runtimes of each operation, and pauses collocation when a foreground job runs an operator that has been observed to suffer large slowdowns.

\section{Implementation}

We implement \sys on the C++ frontend of a recent release of PyTorch (v1.11.0 \cite{libtorch}) using CUDA (v11.6 \cite{cuda}) and NVIDIA's Collective Communications Library (NCCL, v2.11.4 \cite{nccl}). Recent additions to both PyTorch and NCCL allow us to easily create CUDA graphs containing full training iterations using CUDA's stream capture function. This function records all operations enqueued to the device during some observation period and enables the host to relaunch all the same operations with a single graph launch.

\section{Evaluation}

\begin{figure*}[t]
\centering
\includegraphics[width=\textwidth]{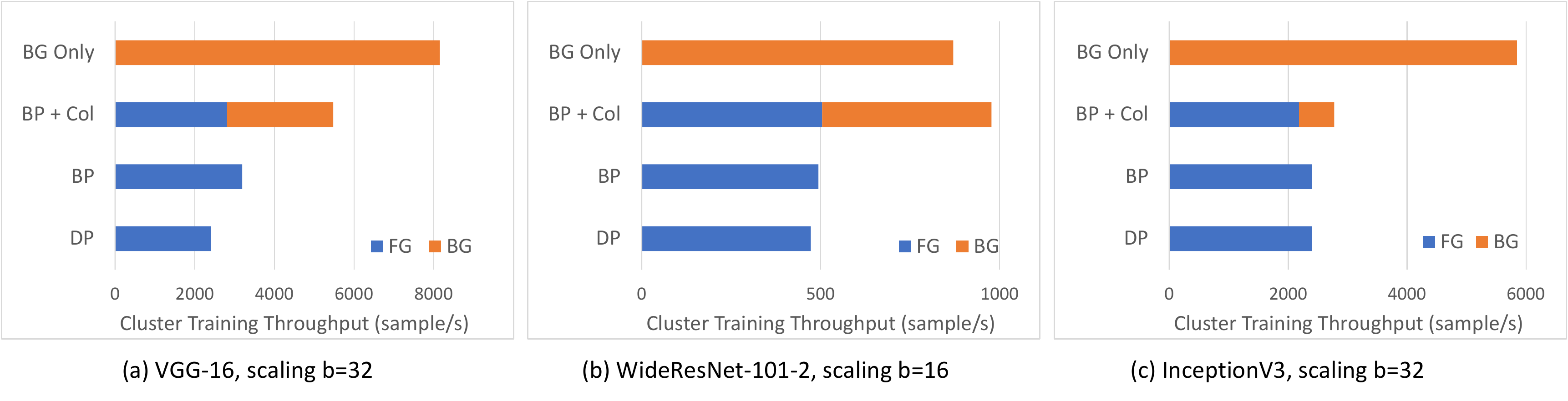}
\vspace{-2em}
\caption{Cluster training throughput while strong scaling with three different scenarios on 8x A100 GPUs. ``BG Only'' shows the maximum throughput achievable when each GPU just runs the background task used for collocation. (a) VGG16: scaled with global batch size of 32. (b) WideResNet-101-2: scaled with global batch size of 16. (c) Inception-V3: scaled with global batch=32. In all three scenarios, using burst parallel scheduling with collocation dramatically improves cluster throughput with little impact on foreground training time.
}
\label{fig:overallResult}
\end{figure*}

Our evaluation aims to answer the following questions:
\begin{enumerate}[itemsep=0mm]
    \item Can we improve the training throughput of each GPU while strong scaling a foreground job?
    \item Does burst parallel training offer better combinations of total cluster throughput and foreground speedup than statically partitioning a cluster?
    \item How do the individual techniques of \sys enable low-interference collocation?
    \item When can we collocate background jobs without interfering with foreground performance?
    \item How fast does DeepPool's burst parallel training planner find the optimized training plan?
\end{enumerate}

\begin{table}
\caption{Workload Characteristics}
\label{tbl:models}
\vskip 0.1in
\centering
\begin{small}
\resizebox{\columnwidth}{!}{\begin{tabular}{lrccc}
\toprule
Model           & Params & Layers & Input Size & Structure\\
\midrule
VGG-16          & 132 M & 21  & 3 x 224 x 224   & Conv, Dense\\
WideResNet-101-2& 127 M & 105 & 3 x 400 x 400  & Intense Conv\\
Inception-v3    & 24 M  & 119 & 3 x 299 x 299  & Light Conv\\
\bottomrule
\end{tabular}}
\end{small}
\vskip -0.1in
\end{table}

To answer the questions above, we measured performance with three vision models: VGG-16~\cite{vgg}, WideResNet-101-2~\cite{zagoruyko2017wide}, and Inception-V3~\cite{inception}. 
Table~\ref{tbl:models} shows the number of operators, parameter size, input size, and main operator type for each network.

We compare the performance of three different scenarios. ``DP'' is our baseline which runs only one data-parallel foreground task by evenly splitting the given global batch across all GPUs.
``BP'' uses burst parallel scheduling (\S\ref{sec:burstParallel}) to scale down stages with low scalability in the foreground task. ``BP+Col'' collocates a low priority background task with the burst-parallel foreground job. Foreground and background tasks use the same model for ease of understanding GPU throughput. We optimized the baseline ``DP'' performance by removing kernel launch overheads with CUDA graphs. This optimization improved the baseline performance up to $2\times$ for some models, and made it more challenging to collocate a background job because graphs enable better utilization of the GPU. To our knowledge, no prior work that has focused on collocating multiple DNN training tasks has made use of this feature.

\begin{table}[]
\caption{Hardware configuration.}
\label{tbl:hardware}
\vskip 0.10in
\centering
\begin{small}
\begin{tabular}{r|l}
\toprule
GPU     & 8 $\times$ NVIDIA A100-SXM4-40GB \\
Interconnect &  NVSwitch (600 GB/s for each GPU) \\
Driver  & CUDA 11.6, cuDNN: v8.3.2, NCCL 2.11.4 \\
CPU     & 2 $\times$ AMD EYPC 64 cores @ 1.5 GHz \\
RAM     & 988 GB \\
OS      & Ubuntu 20.04 (Linux 5.13.0-28-generic) \\
\bottomrule
\end{tabular}
\end{small}
\vskip -0.1in
\end{table}

Table~\ref{tbl:hardware} shows the hardware configurations for our experiments. Although our work is targeting very large scaling with hundreds of GPUs, we use a single NVIDIA DGX A100 server with 8 GPUs. For the purpose of demonstrating improvements on strong scaling efficiency, we use small per-GPU batch sizes which are required for large-scale training (as seen in \S\ref{sec:challenge}) for our 8-GPU experiments, and enable Automatic Mixed Precision in PyTorch to leverage faster implementations of certain operators.

\begin{figure*}
\centering
\includegraphics[width=\textwidth]{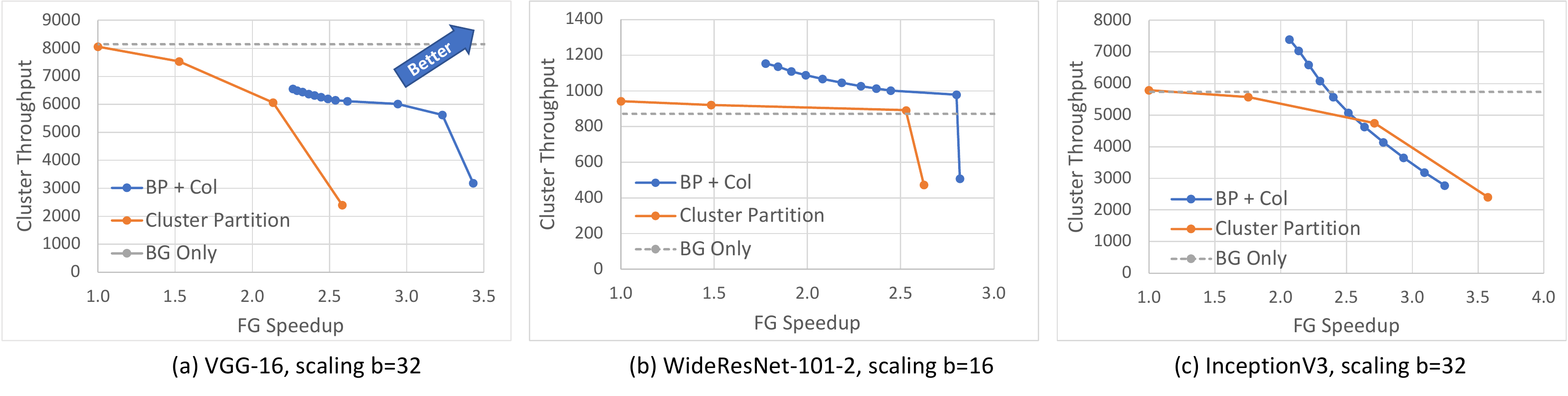}
\vspace{-2em}
\caption{Trade-offs between cluster total throughput and foreground job speedup with various operating points of ``BP + Col'' scenario for the workloads in Figure~\ref{fig:overallResult}. Speedup is calculated relative to the same job running on a single GPU with the same global batch size. As the baseline, ``Cluster Partition'' shows four static partition configurations: 1, 2, 4, 8 GPUs for data-parallel foreground task and 7, 6, 4, 0 GPUs for background tasks respectively.}
\label{fig:paretoFront}
\end{figure*}

\subsection{Overall performance}

To demonstrate that \sys can achieve good training throughput while strong scaling a foreground task, Figure~\ref{fig:overallResult} compares the cluster-wide training throughput of foreground and background tasks with three different scenarios. For the two scenarios that use burst parallelism, we set the GPU-sec amplification limit and collocation parameters to minimize the impact on the foreground performance while having a reasonable gain on total training throughput. 
For VGG-16 and WideResNet-101-2, burst parallel scheduling alone (BP) improves foreground throughput compared to DP as it scales down the number of GPUs used for layers with low scalability, reducing gradient sync overheads. Compared to ``BP'', ``BP + Col'' almost doubles the total cluster throughput with less than 18\% degradation of foreground throughput, which is still higher than DP with 8 GPUs.
For Inception-v3, the burst parallel schedule falls back to standard data-parallel strong scaling since the layers in Inception-v3 have largely even scalability across stages. ``BP + Col'' only extracts an additional 15\% cluster throughput through collocation. This is because Inception-v3 contains a large number of layers with very short kernel execution times on the GPU, which makes it more vulnerable to interference from a background task. We explore this issue in more detail in \S\ref{subsec:collocatability}. Additionally, without burst parallel scaling, there are fewer resources left for the background job to use.

To demonstrate \sys{}'s benefit in depth, Figure~\ref{fig:paretoFront} compares foreground speedups and cluster-wide training throughput for two different scenarios. 
It shows multiple operating points of ``BP + Col'' by varying the GPU-sec amplification limit and collocation parameters. As the baseline, it also shows multiple ``Cluster Partition'' configurations, which divide the 8 GPUs into data-parallel training groups for the foreground task and individual local training groups for the background tasks. For VGG-16, ``BP + Col'' configurations can achieve up to 38\% higher foreground speedups than the ``Cluster Partition'' configurations that provides the same cluster throughput. For WideResNet-101-2, ``BP + Col'' provides not only better foreground speedup (up to 11\%) but also higher cluster throughput (up to 25\%). For Inception-V3, ``BP + Col'' allows up to 37\% higher foreground speedup when we want to operate the cluster with higher throughput than 2 foreground GPU cluster partitions.

\subsection{Decomposing benefits of each technique}
\begin{figure}[t]
\centering
\includegraphics[width=\columnwidth]{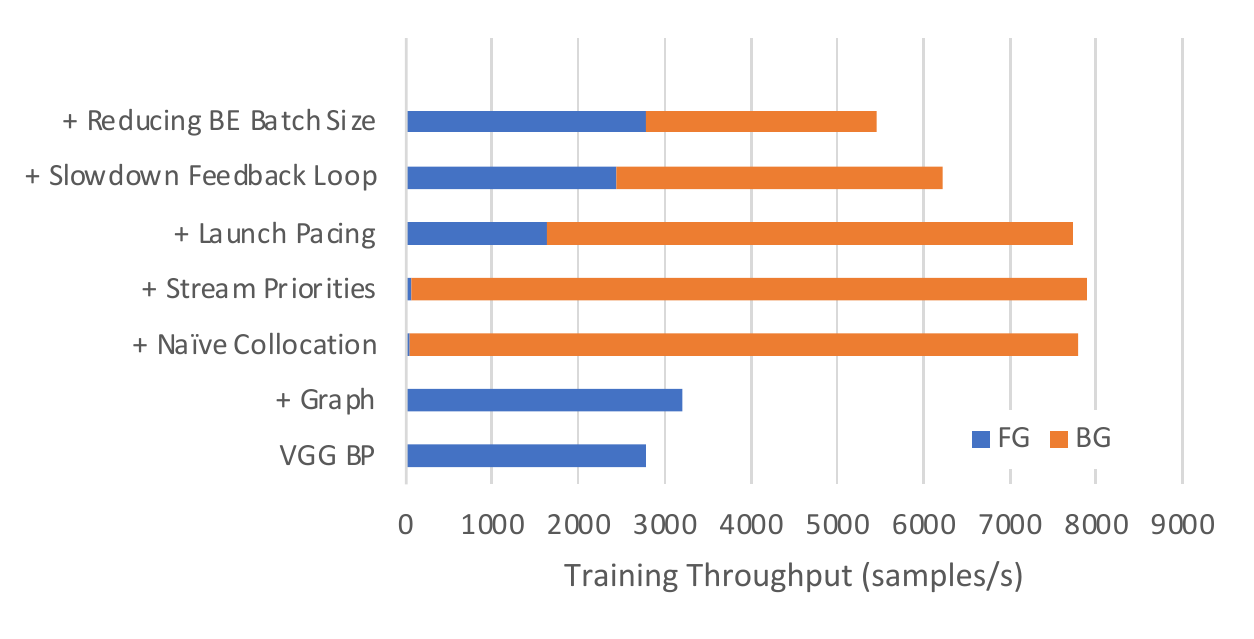}
\vspace{-2em}
\caption{The contribution of each individual multiplexing mechanism towards QoS and throughput when multiplexing VGG-16 on a cluster with 8x A100 GPUs. From the bottom row up, each row illustrates the impact of the addition of each technique. }
\vspace{-1em}
\label{fig:multiplex}
\end{figure}

To illustrate the contributions of each mechanism for multiplexing in \sys discussed in \S\ref{sec:multiplexing}, we iteratively add each mechanism and show its impact on system throughput and QoS for VGG-16 in Figure~\ref{fig:multiplex}. Adding CUDA graphs yields a $15\%$ gain in throughput. Models with larger numbers of small kernels may benefit even more. For example, Inception-V3 sees a $2.2\times$ speedup when enabling CUDA graphs. Naively running a second lower-priority instance of VGG-16 dramatically reduces the throughput of the main task, and adding the use of stream priorities to differentiate between the two has little impact. By limiting the number of outstanding operations enqueued to the device by each task (``launch pacing''), \sys{} reduces contention in shared device queues and improves the throughput for the high-priority task, allowing stream priorities to be used by the device to more effectively differentiate between high and low-priority tasks. By monitoring the slowdown of each individual operator and banning collocation for highly sensitive operators (``slowdown feedback loop''), \sys{} further improves the QoS for the high-priority task. Finally, \sys{} reduces the batch size of the background job in order to reduce the impact of having a non-preemptive scheduler on the foreground job. 

\subsection{Analysis on collocatability}
\label{subsec:collocatability}
\begin{figure}[t]
\centering
\includegraphics[width=\columnwidth]{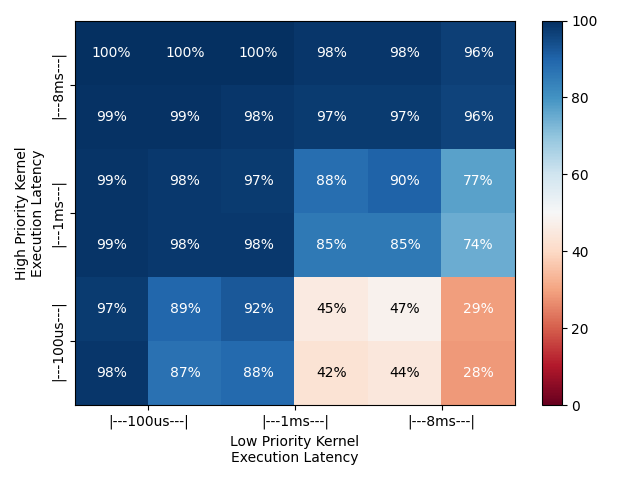}
\vspace{-2em}
\caption{Pairwise collocation of various synthetic CUDA kernels with varied compute intensity and execution latency using stream priorities. Each box shows the high priority kernel's achieved throughput in each collocation as a percentage of its peak achievable throughput when run in isolation. CUDA stream priorities are largely effective, but the non-preemptive device scheduler struggles to provide good QoS for short kernels when collocated with low-priority long-running kernels. }
\label{fig:pairwisekernels}
\end{figure}

To better understand the limits of multiplexing, we run a microbenchmark to explore the impact of non-preemption in the GPU scheduler. We examine the pairwise collocation of several synthetic kernels with varied compute intensities and execution latencies. Figure~\ref{fig:pairwisekernels} shows the throughput degradation for each kernel when run in a high-priority stream when each other kernel is run in a low-priority stream. We observe that streams priorities are largely effective except for in the case where the high-priority tasks have very low execution latency while the low-priority tasks have high execution latency. For this reason, \sys limits the batch size used in best-effort tasks, resulting in lower latency kernels that cause less interference. Preemptive scheduling on the device would allow for better QoS for short-running high-priority kernels.

\subsection{Performance of burst parallel training plan search}

DeepPool's burst parallel training planner uses dynamic programming and graph reduction to find the optimal parallelization plan. To prove that these two search methods are fast enough to be practical, we benchmarked our single-threaded Python implementation of the search algorithm. 

\begin{table}
\caption{Time to search for burst parallel training plans}
\label{tbl:searchTime}
\vskip 0.1in
\centering
\begin{small}
\begin{tabular}{lrr}
\toprule
Scale           & 8 GPUs & 1024 GPUs \\
\midrule
VGG-16          & 0.01 sec & 0.05 sec  \\
WideResNet-101-2& 0.02 sec & 0.11 sec  \\
Inception-v3    & 0.22 sec & 3.23 sec  \\
\bottomrule
\end{tabular}
\end{small}
\vskip -0.1in
\end{table}

Table~\ref{tbl:searchTime} shows the measured search time for two settings: 8 GPUs and 1024 GPUs. For all models, search completes within a few seconds. To limit the growth of the search space, our planner only considers GPU counts that are powers of two. Thanks to this optimization, search time grows only by 5-15$\times$ even when we scale from 8 GPUs to 1024 GPUs. The search for Inception-v3 takes longer than other models since it has multiple branches, so its search must perform the graph reduction algorithm. 

\section{Related Work}

\textbf{Statistical efficiency of training.} Increasing the mini-batch size to scale out the number of GPUs often reduces the statistical efficiency of training, resulting in a much longer training time to reach the same accuracy. A wide study ~\cite{shallue2018measuring} showed that the statistical efficiency depends on model architecture and data set. There were efforts to increase the maximum sizes of mini-batches without hurting the statistical efficiency~\cite{goyal2017accurate, mikami2018massively, you2017large}. However, those works were often targeted to specific DNN architectures, so it's not obvious how to apply the same technique to other new architectures. Statistical efficiency is now considered for cluster schedule optimization as well. Pollux~\cite{pollux} is a cluster scheduler that measures the change of statistical efficiency and uses the effective maximum mini-batch size to optimize  training goodput.

\textbf{Collocating multiple jobs on a GPU.} Several prior works have observed low utilization of both compute and memory GPU resources during DNN training and propose concurrent execution of either multiple training jobs or parallel stages within a single training task to improve utilization. 
Gandiva~\cite{xiao2018gandiva} attempts to pack multiple jobs onto a single GPU when the cluster is overloaded, and avoids colocations that result in a net throughput loss. Wavelet~\cite{wavelet} and Salus~\cite{MLSYS2020_f7177163} aim to improve GPU utilization by interleaving or overlapping execution stages that have different levels of memory usage. 
IOS ~\cite{ios} and Rammer~\cite{rammer} exploit inter-operator parallelism to maximize job throughput, concurrently running operators from parallel branches of a single training task. 
While several of these works identify host-side scheduling overheads as a contributing factor to GPU compute underutilization, none leverage CUDA graphs to amortize kernel launch costs.
Additionally, none of these works aim to provide performance isolation between concurrently executing foreground and background jobs.

\textbf{Strong scaling on other dimensions.} Another approach to tackle the diminishing speedup from strong scaling is scaling on dimensions beyond samples (i.e., reducing samples per GPU). Recent work~\cite{jia2018data, granularStrongScale} explored other dimensions such as splitting height/width of image samples or splitting parameters. However, those unconventional splits incur communication overhead after every layer to prepare input for the next layer (e.g., halo exchange). Our parallelization strategy planner can explore those other dimensions for strong scaling. However, with the latest hardware and new vendor-provided tools like CUDA graphs, the computation cost of each layer is often reduced below 100 us, so the communication penalty of those unconventional splits now exceeds the improvement on computation time. Thus, we primarily focused on burst scaling on the sample dimension only.
Pipeline parallelism~\cite{pipedream, gpipe} is often used to scale training throughput. But pipeline parallelism doesn't directly reduce iteration time and sacrifices statistical efficiency, so we don't consider it in this work.

\textbf{Gradient all-reduce overhead.} As we reduce the iteration time through strong scaling, the gradient all-reduce overhead becomes more critical. Along with improvements in networking speed, there were many efforts on reducing the overhead. To reduce the actual number of bytes sent over the network, prior works have investigated various forms of gradient compression, notably, quantization \cite{seide20141, alistarh2016qsgd} and sparsification \cite{strom2015scalable, dryden2016communication, lin2017deep}, and also have designed efficient collective operations tailored for inherently sparse data \cite{renggli2019sparcml, fei2021efficient}.
Leveraging programmable switch hardware capabilities, the idea of in-network aggregation has been revisited in the context of deep neural network training \cite{sapio2019scaling, lao2021atp}.

\textbf{Cluster-wide optimization.} As the compute demands for deep neural network training have grown steeply, it has become the norm to build a cluster with concentrated compute resources and let users share them. To efficiently serve multiple training jobs being submitted to the shared cluster in real-time, various dynamic scheduling algorithms have been developed targeting a wide range of performance metrics such as average job completion time \cite{peng2018optimus, gu2019tiresias, shin2021elastic}, fairness \cite{mahajan2020themis}, cluster utilization \cite{xiao2018gandiva}, and throughput-cost efficiency~\cite{MLSYS2020_006f52e9}.

\section{Conclusion}

We have introduced two techniques to improve the efficiency of strong scaling: burst parallel training and GPU multiplexing. By bursting the number of GPUs only to stages that can benefit from the additional GPUs, we save GPU resources. We showed that the saved resources can be reclaimed by a collocated background job with only a small impact to the foreground job. We also show that further improvements to hardware design can yield efficiency gains.
These two techniques significantly improve the overall cluster throughput while strong scaling a time-critical foreground job. 

\section*{Acknowledgments}

We thank our anonymous MLSys reviewers for their feedback. This work was funded by the DARPA FastNICs program under contract \#HR0011-20-C-0089.

\clearpage

\bibliography{main}
\bibliographystyle{mlsys2022}


\end{document}